\documentclass[onecolumn,showpacs,amsmath,amssymb]{revtex4-1}

\usepackage[dvipdfmx]{graphicx}

\begin{document}


\title{Neutrino Flavor Ratios Modified by Cosmic Ray Secondary-acceleration}


\author{Norita Kawanaka}
\email{norita@astron.s.u-tokyo.ac.jp}
\affiliation{Department of Astronomy, Graduate School of Science, University of Tokyo, 7-3-1, Hongo, Bunkyo-ku, Tokyo, 113-0033, Japan}
\author{Kunihito Ioka}
\affiliation{Theory Center, Institute of Particle and Nuclear Studies, KEK, Tsukuba 305-0801, Japan}
\affiliation{Department of Particle and Nuclear Physics, SOKENDAI (The Graduate University for Advanced Studies), Tsukuba 305-0801, Japan}

\begin{abstract}
Acceleration of $\pi$'s and $\mu$'s modifies the flavor ratio at Earth (at astrophysical sources) of neutrinos produced by $\pi$ decay, $\nu_e:\nu_\mu:\nu_\tau$, from $1:1:1$ ($1:2:0$) to $1:1.8:1.8$ ($0:1:0$) at high energy, because $\pi$'s decay more than $\mu$'s during secondary-acceleration.  The neutrino spectrum accompanies a flat excess, differently from the case of energy losses.  With the flavor spectra, we can probe timescales of cosmic-ray acceleration and shock dynamics.  We obtain general solutions of convection-diffusion equations and apply to gamma-ray bursts, which may have the flavor modification at around PeV -- EeV detectable by IceCube and next-generation experiments.
\end{abstract}

\date{\today}
\maketitle

\section{Introduction}
The origin of high energy cosmic rays (CRs) has been a long-standing problem in astrophysics.  Especially, CRs with energy above $\gtrsim 10^{19}~{\rm eV}$ are considered to come from extra-Galactic sources such as active galactic nuclei (AGNs) and gamma-ray bursts (GRBs).  In these sources we expect the production of high energy neutrinos ($\gtrsim 0.1~{\rm TeV}$) through interactions of accelerated protons with the ambient photons ($p\gamma$ interactions) or gas ($pp$ or $pn$ interactions) \cite{waxmanbahcall97, murase+06, muraseioka13}.  Detection of these neutrinos can provide us new information about high energy cosmic ray sources as well as the acceleration processes.

In cosmic ray accelerators, high energy neutrinos are mainly produced from the decay of charged pions: $\pi^+ \rightarrow \mu^+ + \nu_{\mu} \rightarrow e^+ +\nu_{\mu} +\nu_e +\bar{\nu}_{\mu}$ and $\pi^- \rightarrow \mu^- + \bar{\nu} \rightarrow e^- + \nu_{\mu} +\bar{\nu}_e + \bar{\nu}_{\mu}$.  Therefore, the flavor ratios of these neutrinos are expected to be
\begin{eqnarray}
\Phi_{\nu_e}^0:\Phi_{\nu_{\mu}}^0:\Phi_{\nu_{\tau}}^0=1:2:0,
\end{eqnarray}
at the sources, where $\Phi_{\nu_i}^0$ denotes the flux of $\nu_i$ and $\bar{\nu}_i$ ($i=e$, $\mu$ or $\tau$).  The observed flavor ratios become $\Phi_{\nu_e}:\Phi_{\nu_{\mu}}:\Phi_{\nu_{\tau}}=1:1:1$ after the neutrino oscillations during the propagation to the Earth \cite{learnedpakvasa95}.  However, this argument may be too naive because we should take into account the finiteness of the decay timescale of pions $\pi^{\pm}$ and muons $\mu^{\pm}$.  For example, if the cooling timescale \citep{kashtiwaxman05, lipari+07, tamborraando15} or acceleration timescale \citep{murase+12, klein+13, winter+14, reynoso14} of a pion or a muon is shorter than the decay timescale, the spectral shape of neutrinos produced from the decay of those particles would be significantly modified.  Especially, because the decay times are different between pions and muons, the energy dependence of neutrino fluxes would be different from flavor to flavor.  The observed neutrino flavor ratio may be also modified by neutron decay \cite{anchordoqui+04} and new physics such as neutrino decay \cite{beacom+03, baerwald+12}, sterile neutrinos \cite{athar+00}, pseudo-Dirac neutrinos \cite{beacom+04, esmaili10}, Lorentz or {\it CPT} violation \cite{hooper+05}, quantum-gravity \cite{anchordoqui+05} and secret interactions of neutrinos \cite{iokamurase14}.

Recently the high energy neutrino detector, IceCube, has discovered $30~{\rm TeV}-2~{\rm PeV}$ neutrinos \cite{aartsen+14} which are confirmed to be non-atmospheric at the level of $5.7\sigma$.  The IceCube team has also analyzed the flavor composition of astrophysical neutrinos in the energy range of $35~{\rm TeV}-1.9~{\rm PeV}$ and demonstrate consistency with $\Phi_{\nu_e}:\Phi_{\nu_{\mu}}:\Phi_{\nu_{\tau}}=1:1:1$ (although the best fit composition is $0:0.2:0.8$).  In the near future, the next generation IceCube-Gen2 and KM3Net experiments will enable the precise study of the energy spectrum of high energy neutrinos and their flavor composition.  

There is no study about how the flavor ratio of observed neutrinos as well as its energy dependence are modified by the acceleration of pions and muons, although several authors have investigated the neutrino energy spectrum under the secondary-acceleration \cite{murase+12, klein+13, winter+14, reynoso14}.  In addition, their approaches are based on one-zone \cite{klein+13, winter+14} or two-zone \cite{reynoso14} approximations, that is, they do not consider the spatial distribution of secondary particles (pions and muons) and their transport across the shock.

In this work, we investigate the acceleration of pions and muons produced by protons by solving their convection-diffusion equations around the shock front taking into account their decay into other particles (i.e., pions into muons and muon neutrinos, and muons into muon neutrinos and electron neutrinos).  The shock acceleration of secondary particles has been discussed in the context of the positron excess \cite{blasi09, blasiserpico09, mertschsarkar09, ahlers+09} observed by PAMELA/Fermi LAT/AMS-02 \citep{adriani+11, ackermann+12, aguilar+13} (see also \citep{ioka10, kawanaka+10, kawanaka12}).  We develop their formalism by including the decay of secondary particles during their acceleration, and evaluate the energy spectrum of neutrinos produced from those particles.

This paper is organized as follows.  In Section 2 we formulate a general model for the shock acceleration of pions and muons, which are produced from shock-accelerated protons via photomeson interactions, and give versatile expressions of the neutrino spectra for later application.  In Section 3, we show that the acceleration of pions/muons would be possible in low-power GRBs occurring inside their progenitors, and apply our model to that case to compute the neutrino spectra and flavor ratios.  Our results and discussions are summarized in Section 4.  In Appendix A, we summarize general solutions of the convection-diffusion equations for pions (decaying secondary particles) and in Appendix B for muons (decaying tertiary particles).

\section{Model}

In this section, we describe the shock acceleration of secondary pions and muons that are generated from the protons accelerated at the shock, and investigate the energy spectrum of neutrinos produced by those pions and muons without specifying a particular source.  Hereafter we neglect the energy loss of particles due to synchrotron emission and inverse Compton scattering for simplicity (see discussions in Sec. 4).

In the shock rest frame, the transport and shock acceleration of particles decaying into other kinds of particles with a timescale $\tau_i$ can be described by the convection-diffusion equation,
\begin{eqnarray}
u\frac{\partial f_i}{\partial x}&=&\frac{\partial}{\partial x}\left[ D(p)\frac{\partial f_i}{\partial x} \right] +\frac{p}{3}\frac{du}{dx}\frac{\partial f_i}{\partial p}-\frac{f_i}{\tau_i}+Q_i(x,p), \label{transport}
\end{eqnarray}
where $f_i(x,p)$ ($i=\pi, \mu$) is the equilibrium distribution function of accelerated particles per unit spatial volume and per unit volume in momentum space, $D(p)$ is the diffusion coefficient, and $u$ is the the velocity of the background fluid.  We assume that the shock is non-relativistic and that the distribution functions are stationary and isotropic except for the shock front for simplicity.  When the shock is relativistic, we should solve the relativistic version of the convection-diffusion equation taking into account the anisotropy of the particle momentum distribution.  The anisotropy is the order of $\beta_-$, and therefore in the mildly-relativistic shock, it would be less than a factor of two.  

The shock front is set at $x=0$, and the upstream (downstream) region corresponds to $x<0$ ($x>0$).  If we ignore the third term of the right-hand side, which describes the loss of particles due to their decay, this is a well-known equation for the usual diffusive shock acceleration of particles \cite{blandfordeichler87, malkovdrury01}.  The decay timescales of a pion and a muon are the functions of their energy,
\begin{eqnarray}
\tau_{\pi}&=&\tau_{\pi,0}\frac{\varepsilon_{\pi}}{m_{\pi}c^2} \nonumber \\
&\simeq& 1.9\times 10^{-2}~{\rm s}~\varepsilon_{\pi, 100{\rm TeV}}, \label{pionlife} \\ 
\tau_{\mu}&=&\tau_{\mu,0}\frac{\varepsilon_{\mu}}{m_{\mu}c^2} \nonumber \\
&\simeq& 2.1~{\rm s}~\varepsilon_{\mu, 100{\rm TeV}}, \label{muonlife}
\end{eqnarray}
where $\varepsilon_i=100~{\rm TeV}\varepsilon_{i,100{\rm TeV}}$ is the energy of a particle at the shock rest frame and, $\tau_{\pi,0}\simeq 2.6\times 10^{-8}~{\rm s}$ and $\tau_{\mu,0}\simeq 2.2\times 10^{-6}~{\rm s}$ are the decay timescales of a pion and a muon at their rest frames, respectively.

The fourth term of the right-hand side of Eq.(\ref{transport}), $Q_i(x,p)$, is the distribution function of $i$ particles injected per unit time.  Here we consider that charged pions are produced in $p\gamma$ interactions, $p\gamma \rightarrow \Delta^+ \rightarrow \pi^+ n$.  In this case, $Q_{\pi}(x,p)$ should be given from the distribution function of primary protons.  On the other hand, in the case of muons, since they are produced by the decay of pions, $Q_{\mu}(x,p)$ should be given from the distribution of pions (see below).

Hereafter, we assume the Bohm-type diffusion,
\begin{eqnarray}
D(p)=\frac{\eta c^2 p}{3eB},
\end{eqnarray}
where $e$ is the charge of a particle, $B$ is the magnetic field strength and $\eta$ is the gyrofactor, which is equal to unity in the Bohm limit \cite{drury83}.  The fluid velocity is given by
\begin{eqnarray}
u(x)=\left \{
\begin{array}{ll}
u_- & (x \leq 0), \\
u_+ & (x>0), \\
\end{array} \right.
\end{eqnarray}
where $u_-$ and $u_+$ are constants and the compression ratio is $\sigma=u_-/u_+>1$.

One should solve Eq.(\ref{transport}) taking the following boundary conditions into account:
\begin{eqnarray}
&{\rm (i)}& \lim_{x \to -0}f_i=\lim_{x \to +0}f_i, \label{cd1} \\
&{\rm (ii)}& \lim_{x \to -\infty}f_i=0,~\lim_{x \to +\infty}f_i<\infty, \label{cd2} \\
&{\rm (iii)}& \left[  D(p)\frac{\partial f_i}{\partial x} \right]_{x=+0}^{x=-0}=\frac{1}{3}(u_+-u_-)p \left. \frac{\partial f_i}{\partial p} \right| _{x=0}, \label{cd3}
\end{eqnarray}
where (iii) comes from the integration of Eq.(\ref{transport}) across the shock front.  This condition yields the differential equation for the distribution function at the shock front $f_{i,0}(p)\equiv f_i(x=0,p)$ with respect of $p$ (see Appendix A).

The detail of the general solution of Eq. (\ref{transport}) is presented in the Appendix A.  In the following subsections we briefly describe the properties of the derived distribution functions of pions and muons.

\subsection{pion acceleration}
In this subsection we show the pion distribution function evaluated from Eq.(\ref{transport}).  Pions are produced through the interactions between the protons accelerated in the shock and the ambient photons.  The distribution of protons is given by
\begin{eqnarray}
f_p(x,p)=\left \{
\begin{array}{ll}
f_{p,0}(p)\exp [ x u_-/D(p) ] & (x\leq 0), \\
f_{p,0}(p) & (x>0), \label{protondis} \\ 
\end{array} \right.
\end{eqnarray}
where $f_{p,0}(p)$ is the proton distribution function at the shock front, which is proportional to $\sim p^{-\gamma}$.  This expression (\ref{protondis}) is a well-known solution of the convection-diffusion equation (\ref{transport}) \cite{blandfordeichler87, malkovdrury01, drury83}.

For simplicity we assume that the ambient photon field is uniform.  Since pions are produced from shock-accelerated protons, the production spectrum at the source of pions $Q_{\pi}$ is proportional to that of primary protons, which can be described as
\begin{eqnarray} 
Q_{\pi}(x,p)=\left \{
\begin{array}{ll}
Q_{\pi,0}(p)\exp \left[ xu_-/D(p_{\rm p}) \right] & (x \leq 0),\\
Q_{\pi,0}(p) & (x>0),\\
\end{array} \right. \label{pioninj}
\end{eqnarray}
where $p_{\rm p}$ is the momentum of a primary proton generating a secondary pion with a momentum $p$, and these momenta are approximately related in a linear way:
\begin{eqnarray}
p\approx \xi_{\pi}p_{\rm p},
\end{eqnarray}
where $\xi_{\pi}\approx 0.2$ is the ratio of the energy of a pion to that of a primary proton \cite{waxmanbahcall97}.

We can solve Eq. (\ref{transport}) for pions by using Eqs. (\ref{solution}), (\ref{gipm}), and (\ref{hipm}) with $Q_{\pi}(x,p)$ in Eq. (\ref{pioninj}), and obtain the pion distribution functions in the upstream $f_{\pi,-}(x,p)$ and downstream $f_{\pi,+}(x,p)$ as 
\begin{eqnarray}
f_{\pi,-}&=&\left[f_{\pi,0}-\frac{DQ_{\pi,0}}{D/\tau_{\pi}+(\xi_{\pi}-\xi_{\pi}^2)u_-^2}\right] \exp\left( \frac{\sqrt{u_-^2+4D/\tau_{\pi}}+u_-}{2D}x \right)+\frac{DQ_{\pi,0}}{D/\tau_{\pi}+(\xi_{\pi}-\xi_{\pi}^2)u_-^2}\exp \left( \frac{\xi_{\pi}u_-}{D}x\right), \label{pionus} \\
f_{\pi,+}&=&\left( f_{\pi,0}-Q_{\pi,0}\tau_{\pi}\right) \exp \left(-\frac{\sqrt{u_+^2+4D/\tau_{\pi}}-u_+}{2D}x \right) +Q_{\pi,0}\tau_{\pi}. \label{pionds}
\end{eqnarray}
The pion distribution functions at the shock front $f_{\pi,0}\equiv f_{\pi}(x=0,p)$ can be evaluated from Eqs. (\ref{ai}), (\ref{gip}) and (\ref{fizeroint}) as
\begin{eqnarray}
f_{\pi,0}(p)=\gamma B_{\pi}\int_0^p \frac{dp^{\prime}}{p^{\prime}}\left( \frac{p^{\prime}}{p} \right)^{\gamma A_{\pi}}\frac{D(p^{\prime})Q_{\pi,0}(p^{\prime})}{u_-^2}, \label{fpizero}
\end{eqnarray}
where $A_{\pi}$ and $B_{\pi}$ are numerical factors, being independent of $p$ (since both $D$ and $\tau_{\pi}$ are proportional to $p$):
\begin{eqnarray}
A_{\pi}&=&\frac{1}{2}\left[ \left( \sqrt{1+\frac{4D}{\tau_{\pi} u_-^2}}+1\right)+\left( \sqrt{\frac{1}{\sigma^2}+\frac{4D}{\tau_{\pi} u_-^2}}-\frac{1}{\sigma} \right) \right], \\
B_{\pi}&=&\frac{2}{\sqrt{1+4D/\tau_{\pi}u_-^2}-(1-2\xi_{\pi})} +\frac{2\sigma}{\sqrt{1+4D/\tau_{\pi}u_+^2}+1},
\end{eqnarray}
where $\sigma=u_-/u_+$ is the compression ratio.  We can see that the distribution function of pions at the shock front, Eq.~(\ref{fpizero}), becomes harder than their production spectrum $Q_{\pi,0}$ by $p^1$ [$\propto D(p)$], and this is similar to Eq.(6) of \cite{blasi09}, where the acceleration of secondary positrons produced in the supernova remnant shock is discussed.  The difference is that we take into account the decay of particles, the third term of the right-hand side of Eq.~(\ref{transport}), in the convection-diffusion equation of secondary particles, which is reflected as numerical factors $A_{\pi}$ and $B_{\pi}$.
 
To see the effects of the transport and acceleration of pions in the downstream region from Eq. (\ref{pionds}), we divide $f_{\pi,+}(x,p)$ into two components: $f_{\pi,{\rm acc}}(x,p)$ and $f_{\pi,{\rm nonacc}}(x,p)$.  The former component $f_{\pi,{\rm acc}}$ represents the pions that are reaccelerated at the shock, being proportional to $D(p)Q_{\pi,0}(p)$.  On the other hand, the latter component $f_{\pi,{\rm nonacc}}$ represents the pions that are produced from the protons and advected in the downstream region, being proportional to $Q_{\pi,0}(p)\tau_{\pi}$:
\begin{eqnarray}
f_{\pi,{\rm acc}}(x,p)&=&f_{\pi,0}(p)\exp \left( -\frac{\sqrt{u_+^2+4D/\tau_{\pi}}-u_+}{2D}x \right),  \label{piacc} \\
f_{\pi,{\rm nonacc}}(x,p)&=&Q_{\pi,0}(p)\tau_{\pi}\left[ 1-\exp \left( -\frac{\sqrt{u_+^2+4D/\tau_{\pi}}-u_+}{2D}x \right) \right]. \label{piadv}
\end{eqnarray}
Hereafter, since the number of upstream pions, $\int_{x<0}dx^3 f_{\pi,-}(x,p)$, is subdominant compared to that of downstream pions, $\int_{x>0}dx^3 f_{\pi,+}(x,p)$, we only discuss the contribution of the downstream pions to the neutrino spectra.
 
In the limit of $D/\tau_{\pi} u_-^2\rightarrow 0$ (i.e., the lifetime of a pion is much longer than the acceleration timescale, $t_{\rm acc}\equiv D/u_-^2$), Eq.(\ref{fpizero}) has the same form as Eq.(6) of \cite{blasi09}.  In this limit, we have $A_{\pi}\approx 1$, $B_{\pi}\approx \xi_{\pi}^{-1}+\sigma$.  Therefore, the distribution function at the shock front $f_{\pi,0}$ in Eq.(\ref{fpizero}) can be approximated as
\begin{eqnarray}
f_{\pi,0}\simeq \frac{\gamma}{\gamma -\alpha +1}\left( \frac{1}{\xi_{\pi}}+ \sigma \right) \frac{D(p)Q_{\pi,0}(p)}{u_-^2} \label{fpizeroapp},
\end{eqnarray}
where $\alpha$ is the power-law index of the production spectrum, $Q_{\pi,0}(p)\propto p^{-\alpha}$, and $\alpha\approx 4$ in the strong shock limit with an adiabatic index $5/3$.  We can see that, since we assume $D(p)\propto p$, the resulting spectrum is proportional to $p^{\alpha +1}$, being harder than the production spectrum.  This can be interpreted as the result of the secondary-acceleration: since pions produced from shock-accelerated protons can cross the shock front before their decay, they would gain the energy and their spectrum would become harder.  We should also note that $f_{\pi,0}$ is proportional to $t_{\rm acc}Q_{\pi,0}$.  In this limit of $D/\tau_{\pi} u_-^2\rightarrow 0$, the pion distribution function in the downstream region (\ref{pionds}) can be approximated as
\begin{eqnarray}
f_{\pi,+}\simeq f_{\pi,0}(p)\exp \left( -\frac{x}{u_+ \tau_{\pi}} \right) +Q_{\pi,0}(p)\tau_{\pi}\left[ 1-\exp \left( -\frac{x}{u_+ \tau_{\pi}} \right) \right],
\end{eqnarray}
where we use $\sqrt{u_+^2+4D/\tau_{\pi}}-u_+\simeq 2D/(u_+ \tau_{\pi})$.  Here the first term corresponds to the pions reaccelerated at the shock, and its damping length scale $u_+ \tau_{\pi}$ is identical to the distance over which a pion is advected with the fluid during its lifetime.  The second term represents the pions that are produced from protons in the downstream region and simply advected further downward.  On the other hand, in the upstream region, Eq.(\ref{pionus}) can be approximated as
\begin{eqnarray}
f_{\pi,-}\simeq \left( f_{\pi,0}(p)-\frac{1}{\xi_{\pi}-\xi_{\pi}^2}\frac{D(p)Q_{\pi,0}(p)}{u_-^2} \right) \exp \left( \frac{u_-}{D}x \right) +\frac{1}{\xi_{\pi}-\xi_{\pi}^2}\frac{D(p)Q_{\pi,0}(p)}{u_-^2} \exp \left( \frac{\xi_{\pi}u_-}{D}x \right),
\end{eqnarray}
where we use $\sqrt{u_-^2+4D/\tau_{\pi}}+u_-\simeq 2u_-$.

\subsection{muon acceleration}

Using the results of the last subsection, we can evaluate the distribution function of muons that are produced from the decay of pions.  The production spectrum at the source of muons $Q_{\mu}$ should be given based on the distribution function of pions as follows:
\begin{eqnarray}
Q_{\mu,\pm}=\frac{f_{\pi,\pm}(p/\xi_{\mu})}{\tau_{\pi}(p/\xi_{\mu})}\frac{dp_{\pi}}{dp}=\frac{1}{\xi_{\mu}}\frac{f_{\pi,\pm}(p/\xi_{\mu})}{\tau_{\pi}(p/\xi_{\mu})}, \label{muoninj}
\end{eqnarray}
where $\xi_{\mu}\approx 0.75$ is the ratio of the momentum of a muon $p$ to that of a primary pion $p_{\pi}$.  Using the method shown in the Appendix A, we can solve the muon convection-diffusion equation and derive $f_{\mu,\pm}(x,p)$.  The detail of the solutions is shown in the Appendix B.  

In the limit of $D/u_-^2 \ll \tau_{\pi},~\tau_{\mu}$ (i.e., the lifetime of a pion is much longer than the acceleration timescale), the muon distribution function at the shock front $f_{\mu,0}(p)$ can be approximated as
\begin{eqnarray}
f_{\mu,0}\simeq \frac{\gamma \xi_{\mu}^{\alpha-1}}{\gamma-\alpha+1}\left[ \left( \frac{1}{\xi_{\mu}}+\sigma \right) \left( \frac{1}{\xi_{\pi}}+\sigma \right) \frac{\gamma}{\gamma-\alpha+1}+\frac{1}{\xi_{\mu} \xi_{\pi}^2} \right] \left( \frac{D(p)}{u_-^2} \right)^2 \frac{1}{\tau_{\pi}} Q_{\pi,0}(p) \label{fmuzeroapp},
\end{eqnarray}
where we assume the power-law spectrum of pion production, $Q_{\pi,0}(p)\propto p^{-\alpha}$.  Since $D(p)\propto p$ and $\tau_{\pi}\propto p$, we can see that this spectrum Eq.(\ref{fmuzeroapp}) is proportional to $p^{\alpha+1}$, which is similar to the pion distribution function at the shock $f_{\pi,0}$ shown in Eq. (\ref{fpizeroapp}).  This can be interpreted as follows. As stated in Eq.(\ref{muoninj}), the production spectrum of muons is proportional to $f_{\pi}/\tau_{\pi}$.  Since the lifetime of a pion $\tau_{\pi}$ is proportional to $p$, the production spectrum at the shock is proportional to $Q_{\mu,0}\simeq f_{\pi,0}/\tau_{\pi}\sim p^{-\alpha+1}/p=p^{-\alpha}$.  Injected muons are reaccelerated at the shock and, according to the similar discussion to that on pions, the muon spectrum at the shock would become harder than their injected spectrum by $p^1$, which comes from the dependence of $D(p)$ on $p$.  We should also note that $f_{\mu,0}$ is proportional to $(t_{\rm acc}/\tau_{\pi})t_{\rm acc}Q_{\pi,0}$.

In a similar way to the pion distribution function, the muon distribution function in the downstream $f_{\mu,+}(x,p)$ can be divided into two components, $f_{\mu,{\rm acc}}(x,p)$ and $f_{\mu,{\rm nonacc}}(x,p)$ as follows:
\begin{eqnarray}
f_{\mu,{\rm acc}}(x,p)&=&f_{\mu,0}\exp \left( -\frac{\sqrt{u_+^2+4D/\tau_{\mu}}-u_+}{2D}x \right), \label{muacc} \\
f_{\mu,{\rm nonacc}}(x,p)&=&f_{\mu,+}(x,p)-f_{\mu,0}\exp \left( -\frac{\sqrt{u_+^2+4D/\tau_{\mu}}-u_+}{2D}x \right), \label{muadv}
\end{eqnarray}
and, as in the case of pions, the number of upstream muons, $\int_{x<0}dx^3f_{\mu,-}(x,p)$ is subdominant compared to that of downstream muons $\int_{x>0}dx^3f_{\mu,+}(x,p)$.

\subsection{neutrino spectra}

From the pion and muon distribution functions calculated above, the neutrino spectrum can be obtained as follows:
\begin{eqnarray}
\Phi_{\nu_{\mu}}^0 (p)&=& \int dx^3 \frac{4\pi p^2}{\xi_{\nu_{\mu}}}\frac{f_{\pi}(x, p/\xi_{\nu_{\mu}})}{\tau_{\pi}(p/\xi_{\nu_{\mu}})},  \label{numu} \\
\Phi_{\bar{\nu}_{\mu}}^0 (p)&=& \int dx^3 \frac{4\pi p^2}{\xi_{\bar{\nu}_{\mu}}} \frac{f_{\mu}(x,p/\xi_{\bar{\nu}_{\mu}})}{\tau_{\mu}(p/\xi_{\bar{\nu}_{\mu}})}, \label{numubar} \\
\Phi_{\nu_e}^0 (p)&=&\int dx^3 \frac{4\pi p^2}{\xi_{\nu_e}}\frac{f_{\mu}(x,p/{\xi_{\nu_e}})}{\tau_{\mu}(p/\xi_{\nu_e})}, \label{nue}
\end{eqnarray}
where $\xi_{\nu_{\mu}}$, $\xi_{\bar{\nu}_{\mu}}$ and $\xi_{\nu_e}$ are the ratios of the energy of a muon neutrino, an anti-muon neutrino, and an electron neutrino to that of their primary particles, respectively.  Since each lepton produced from the decay of a pion ($e$, $\nu_{\mu}$, $\bar{\nu}_{\mu}$ and $\nu_e$) carries approximately equal energy (i.e., 1/4 of that of the primary pion), we set $\xi_{\nu_{\mu}}\approx 0.25$, $\xi_{\bar{\nu}_{\mu}}\approx 0.33$ and $\xi_{\nu_e}\approx 0.33$.  The volume integral should contain the surface area integral on the shocked matter plus the integral along the normal direction of the shock.  Especially, defining the dynamical timescale $t_{\rm dyn}$ as time for the shock to cross the system, the latter integral should be from $x\approx  -\beta_-ct_{\rm dyn}$ to $x\approx \beta_+ct_{\rm dyn}$.

We divide the neutrino energy spectrum into two components according to the decomposition of the pion/muon distribution functions shown in Eqs. (\ref{piacc}), (\ref{piadv}), (\ref{muacc}) and (\ref{muadv}):
\begin{eqnarray}
\Phi_{\nu_{\mu},{\rm acc}}^0 (p)&=& \int dx^3 \frac{4\pi p^2}{\xi_{\nu_{\mu}}}\frac{f_{\pi,{\rm acc}}(x,p/\xi_{\nu_{\mu}})}{\tau_{\pi}(p/\xi_{\nu_{\mu}})}, \\
\Phi_{\nu_{\mu},{\rm nonacc}}^0 (p)&=&\int dx^3 \frac{4\pi p^2}{\xi_{\nu_{\mu}}}\frac{f_{\pi,{\rm nonacc}}(x,p/\xi_{\nu_{\mu}})}{\tau_{\pi}(p/\xi_{\nu_{\mu}})},
\end{eqnarray}
and $\Phi_{\bar{\nu}_{\mu},{\rm acc/nonacc}}^0$ and $\Phi_{\nu_e,{\rm acc/nonacc}}^0$ are defined in similar ways.

We should also consider neutrino oscillations during the propagation from the source to the Earth.  When neutrinos propagate over the distances much longer than $\sim \hbar c \epsilon_{\nu}/\Delta m^2 c^4$ ($\Delta m^2$ is the squared mass difference: $\Delta m_{12}^2\simeq 8.0\times 10^{-5}~{\rm eV}^2$, $|\Delta m_{23}^2|\simeq 2.5\times 10^{-3}~{\rm eV}^2$), the observed fluxes of neutrinos $\Phi_{\nu_x}$ ($x=e, \mu, \tau$) should be described as
\begin{eqnarray}
\Phi_{\nu_x}=\sum_y P_{xy}\Phi_{\nu_y}^0=\sum_y \sum_i \left| U_{xi} \right| ^2 \left| U_{yi} \right| ^2 \Phi_{\nu_y}^0, \label{mixing}
\end{eqnarray}
where $U_{xi}$ is the neutrino mixing matrix and the subscript $i$ represents the mass eigenstate of neutrinos.  The matrix elements of $U_{xi}$ can be described by the mixing angles $\theta_{12}$, $\theta_{23}$, and $\theta_{31}$, and the Dirac phase $\delta$.  Based on \cite{fogli+12}, we adopt $\sin ^2\theta_{12}\simeq 0.31$, $\sin ^2\theta_{23}\simeq 0.39$, $\sin ^2\theta_{31}\simeq 0.024$, and $\delta\simeq 1.1\pi$.

\section{Applications to Low-Power GRBs}
Now we consider long GRBs as neutrino sources.  GRBs are thought to produce high-energy neutrinos \cite{waxmanbahcall97}.  In the standard model, the emission of long GRBs is believed to be produced by relativistic jets launched when a massive star collapses and a stellar-mass black hole is formed.  In order for the jet to be observed as a GRB, it should penetrate the stellar envelope, otherwise the jet would stall inside the star and the gamma-ray emission would not be observed \cite{reesmeszaros94}.  Their prompt emission is often interpreted as synchrotron emission from non-thermal electrons accelerated at internal shocks.  It is natural to consider the proton acceleration and the associated production of high energy neutrinos via $pp$/$p\gamma$ interactions \cite{waxmanbahcall97}.

  However, IceCube gave stringent upper limits on GRBs \cite{abbasi+11, abbasi+12} and has ruled out the typical long GRBs as the main source of the observed diffuse neutrino events \cite{hummer+12, he+12, gao+13}.

Instead of ordinary GRBs, we investigate high-energy neutrino production by low-power GRBs such as low-luminosity GRBs (LLGRBs) and ultralong GRBs (ULGRBs), which are still not strongly constrained by IceCube.  In these low-power GRBs the high energy neutrinos may be produced inside the progenitor star \cite{meszaroswaxman01, razzaque+04, horiuchiando08}.  While a jet is penetrating in a stellar envelope, it becomes slow and cylindrical by passing through the collimation shock.  The internal shocks would also occur when there is spatial inhomogeneity in a jet.  Murase \& Ioka \cite{muraseioka13} recently investigated such high energy neutrino production expected from LLGRBs \cite{soderberg+06} and ULGRBs \cite{gendre+13, levan+14}, which have longer durations ($\sim 10^3-10^4~{\rm s}$) and lower luminosity ($L_{\gamma}\sim 10^{46}-10^{50}~{\rm erg}~{\rm s}^{-1}$) compared to those of typical long GRBs.  It has been suggested that ultra-long GRBs have bigger progenitors like blue supergiants (BSGs) with radii of $\sim 10^{12}-10^{13}~{\rm cm}$ \cite{suwaioka11, kashiyama+13, nakauchi+13}.  We apply our model of neutrino production in such GRB jets inside stars, taking into account the secondary-acceleration and decay of pions/muons that are produced by shock-accelerated protons via $p\gamma$ interactions.  The internal shocks of GRBs are considered to be mildly-relativistic in the shock rest frame.

Let us evaluate the important timescales in our model by considering the internal shock scenario of GRBs.  When two moving shells are ejected with comparable Lorentz factors of order of $\Gamma$ from the central engine during the time separation $\Delta t$, these shells collide and make an internal shock at the radius $r \sim r_{\rm is} \sim \Gamma ^2 c\Delta t$.  Here the magnetic field energy density can be estimated as $U_B\equiv L_B/(4\pi r_{\rm is}^2 c \Gamma^2)$, where $L_B$ is the magnetic luminosity.  Then we can estimate the acceleration timescale $t_{\rm acc}$, synchrotron cooling timescale $t_{i,{\rm syn}}$ as functions of the energy of a particle, and the dynamical timescale $t_{\rm dyn}$ at the shock rest frame as follows:

\begin{eqnarray}
t_{\rm acc}&=&\frac{D(p)}{u_-^2}=\frac{\eta \varepsilon_i}{3ceB \beta_-^2} \nonumber \\
&\simeq &4.4\times 10^{-5}~{\rm s}~\frac{\eta \varepsilon_{i, 100{\rm TeV}} \Gamma_2^3 \Delta t_{\rm ms}}{L_{B,47}^{1/2} \beta_-^2}, \\
t_{\pi, {\rm syn}}&=&\frac{9m_{\pi}^4 c^7}{4e^4 B^2 \varepsilon_{\pi}} \nonumber \\
&\simeq&3.0~{\rm s}~\frac{\Gamma_2^6 \Delta t_{\rm ms}^2}{L_{B,47}\varepsilon_{\pi,100{\rm TeV}}}, \\
t_{\mu, {\rm syn}}&=&\frac{9m_{\mu}^4 c^7}{4e^4 B^2 \varepsilon_{\mu}} \nonumber \\
&\simeq& 0.99~{\rm s}~\frac{\Gamma_2^6 \Delta t_{\rm ms}^2}{L_{B,47}\varepsilon_{\mu,100{\rm TeV}}}, \\
t_{\rm dyn}&=&\frac{r_{\rm is}}{\beta_- c\Gamma} \nonumber \\
&=&0.10~{\rm s}~\Gamma_2\Delta t_{\rm ms}\beta_-^{-1},
\end{eqnarray}
where $\varepsilon_i=100~{\rm TeV}~\varepsilon_{i, 100~{\rm TeV}}$ is the energy of a particle $i$ ($i=\pi$ or $\mu$) at the shock rest frame, $\Gamma_2=\Gamma/10^2$, $\Delta t_{\rm ms}=\Delta t/(10^{-3}~{\rm s})$, $L_{47}=L_B/10^{47}~{\rm erg}~{\rm s}^{-1}$, $m_{\pi}\approx 140~{\rm MeV}$ and $m_{\mu}\approx 106~{\rm MeV}$ are the masses of a charged pion and a muon, respectively.

From Eqs. (\ref{pionlife}) and (\ref{muonlife}), a pion can be accelerated at the source before its decay when $t_{\rm acc}<\tau_{\pi}$, i.e.,
\begin{eqnarray}
\frac{\eta\Gamma_2^3 \Delta t_{\rm ms}}{L_{B,47}^{1/2}\beta_-^2}\lesssim 4.4\times 10^2,
\end{eqnarray}
while a muon can be accelerated before its decay when
\begin{eqnarray}
\frac{\eta\Gamma_2^3 \Delta t_{\rm ms}}{L_{B,47}^{1/2}\beta_-^2}\lesssim 4.7\times 10^4.
\end{eqnarray}
Note that, since both $t_{\rm acc}$ and $\tau_{\pi}$ ($\tau_{\mu}$) are proportional to the energy of a pion (a muon), these conditions are independent of the energy of particles.  Under these conditions, pions (muons) can be accelerated at the shock before they decay and therefore their spectra would become harder.  On the other hand, we can see that the synchrotron cooling timescale would be shorter than the acceleration timescale when the energy $\varepsilon_i$ in the shock rest frame is higher than $\varepsilon_{i, 0}$, where
\begin{eqnarray} 
\varepsilon_{\pi,0}\simeq 2.7\times 10^{16}~{\rm eV}\frac{\Gamma_2^{3/2} \Delta t_{\rm ms}^{1/2}\beta_-}{L_{B,47}^{1/4}\eta^{1/2}}, \label{epizero} \\
\varepsilon_{\mu,0}\simeq 1.9\times 10^{16}~{\rm eV}\frac{\Gamma_2^{3/2} \Delta t_{\rm ms}^{1/2}\beta_-}{L_{B,47}^{1/4}\eta^{1/2}}. \label{emuzero}
\end{eqnarray}

In order to evaluate the timescales of inverse Compton cooling and $p\gamma$ interaction, we should give the target photon spectrum at the local rest frame.  In the case of the internal shock occurring inside a star, the accelerated particles mainly interact with photons that are produced in the jet head and escape back from there.  Here we estimate the spectrum of the target photon field according to the procedure adopted in \cite{muraseioka13}.  At the head of the collimated jet, the photon temperature $T_{\rm cj}$ is given as
\begin{eqnarray}
k_{\rm B}T_{\rm cj}&\approx&k_{\rm B}\left( \frac{L}{4\pi r_{\rm cs}^2 \Gamma_{\rm cj}^2 \cdot 4\sigma_{\rm SB}} \right)^{1/4} \nonumber \\
&\approx &0.52~{\rm keV} \epsilon_{B,-2}^{-1}L_{B,47}^{1/4}r_{{\rm cs},11.5}^{-1/2}(\Gamma_{\rm cj}/5)^{-1/2},
\end{eqnarray}
where $L=L_B/\epsilon_B$ is the total jet luminosity ($\epsilon_B=0.01\epsilon_{B,-2}$ is the fraction of the magnetic energy), $k_{\rm B}$ is the Boltzmann constant, $\sigma_{\rm SB}$ is the Stefan-Boltzmann constant, $r_{\rm cs}$ is the radius where a jet becomes cylindrical through the collimation shock, and $\Gamma_{\rm cj}$ is the Lorentz factor of the collimated jet (note that this is different from the Lorentz factor of the precollimated jet $\Gamma$).  The fraction of photons escaping the collimated jet is $f_{\rm esc}\approx (n_{\rm cj}\sigma_T r_{\rm cs}/\Gamma_{\rm cj})^{-1}$ where $n_{\rm cj}\approx L/(4\pi r_{\rm cs}^2\Gamma_{\rm cj}\Gamma m_p c^3)$ is the comoving proton number density in the collimated jet, and $\sigma_T$ is the Thomson cross section.  Therefore, the number density of the target photons is given as
\begin{eqnarray}
n_{\gamma}^j&\approx &\frac{\Gamma}{2\Gamma_{\rm cj}}f_{\rm esc}n_{\gamma}^{\rm cj} \nonumber \\
&\approx & 9.8\times 10^{21}~{\rm cm}^{-3}~\epsilon_{B,-2}^{-1}L_{B,47}^{-1/4}r_{{\rm cs},11.5}^{-1/2}\Gamma_2^2 (\Gamma_{\rm cj}/5)^{-1/2},
\end{eqnarray}
where $n_{\gamma}^{\rm cj}=16\pi \zeta(3)(k_{\rm B}T_{\rm cj})^3/(ch)^3$ is the comoving photon number density in the collimated jet, and $\zeta(n)$ is the Riemann zeta function.  We assume that the escaping photon field has a thermal spectrum,
\begin{eqnarray}
\frac{dn}{d\varepsilon}=\frac{8\pi \varepsilon^2}{c^3 h^3}\frac{1}{e^{\varepsilon/k_{\rm B}T_{\rm eff}}-1},
\end{eqnarray}
 with the effective temperature of $k_{\rm B}T_{\rm eff}\approx [(\Gamma/2\Gamma_{\rm cj})f_{\rm esc}]^{1/3}k_{\rm B}T_{\rm cj}$.

The photomeson production ($p\gamma$ interaction) timescale can be evaluated as
\begin{eqnarray}
t_{p\gamma}^{-1}=\frac{c}{2\gamma_p^2}\int_{\varepsilon_0}^{\infty}d\varepsilon \sigma_{\pi}(\varepsilon)\xi(\varepsilon)\varepsilon \int_{\varepsilon/2\gamma_p}^{\infty}dx x^{-2} \frac{dn}{dx},
\end{eqnarray}
where $\gamma_p=\varepsilon_p/(m_p c^2)$, $\sigma_{\pi}(\varepsilon)$ is the cross section of pion production as a function of photon energy $\varepsilon$ in the proton rest frame, $\xi(\varepsilon)$ is the average fraction of energy lost from a proton to a pion, and $\varepsilon_0=0.15~{\rm GeV}$ is the threshold energy \citep{waxmanbahcall97}.  In the following discussion, we use the $\Delta$ resonance approximation: $\sigma_{\pi}(\varepsilon)$ is approximated to be a function with a peak at $\varepsilon =\varepsilon_{\rm peak}\sim 0.3~{\rm GeV}$, where $\sigma(\varepsilon_{\rm peak})\simeq 5\times 10^{-28}~{\rm cm}^2$ with the width of $\Delta \varepsilon\simeq 0.2~{\rm GeV}$, and $\xi(\varepsilon_{\rm peak})\equiv \xi_{\pi}\simeq 0.2$.

Figures 1 and 2 depict the acceleration timescales, cooling timescales via synchrotron emission and inverse Compton scattering, decay timescales of a pion and a muon, the timescale of $p\gamma$ interactions, and the dynamical timescale ($\equiv r_d/\beta_- c\Gamma$) in the internal shock occurring inside a star expected for an ultralong GRB ($L=10^{49}~{\rm erg}~{\rm s}^{-1}$, $\epsilon_B=0.01$, $\Gamma=80$, $\Delta t=10^{-3}~{\rm s}$, $\beta_-=0.5$).  We can see that, with the current choice of parameters, the acceleration timescales of a pion and a muon are shorter than their lifetimes for arbitrary energy range, and that the decay timescale becomes longer than the dynamical timescale above the energy of $\sim {\rm PeV}$ for pions and $\sim 10~{\rm TeV}$ for muons (at the shock rest frame).  Note that, since synchrotron cooling timescale for pions and muons becomes shorter than acceleration timescale when the energy of particles is larger than $\sim 10~{\rm PeV}$, our formalism is not applicable in the energy range above $\sim 10~{\rm PeV}$.  Note also that the efficiencies of pion/muon production would be suppressed in the energy range where the timescale of $p\gamma$ interactions is comparable ($10^{14}~{\rm eV}\lesssim \varepsilon_i \lesssim 10^{15}~{\rm eV}$) or shorter than the acceleration timescale.  In the current work, this effect is not taken into account.

Figure 3 depicts the energy spectra of muon neutrinos and electron neutrinos expected from the internal shock inside a progenitor of an ultralong GRB.  The flux from reaccelerated pions and muons, Eqs. (\ref{piacc}) and (\ref{muacc}), and that from advected pions and muons, Eqs. (\ref{piadv}) and (\ref{muadv}), are also shown (with dotted lines and dashed lines, respectively).  We can see that the electron neutrino flux from advected muons drops above the energy of $\sim 100~{\rm TeV}$ in the observer frame ($\sim {\rm TeV}$ at the shock rest frame).  This corresponds to the energy at which the muon decay timescale is equal to the dynamical timescale.  Above this energy, only part of muons can decay into $\nu_e$ within the dynamical timescale \cite{muonadcool}.  As for the muon neutrino flux from advected particles, it slightly drops around the energy where the electron neutrino flux drops because anti-muon neutrinos $\bar{\nu}_{\mu}$ are generated from the decay of muons $\mu^+$, and drops again at the energy where the pion decay timescale is equal to the dynamical timescale ($\varepsilon_{\pi}\sim 0.1~{\rm PeV}$ for the current parameter set) because muon neutrinos $\nu_{\mu}$ are generated from the decay of pions $\pi^+$.  We can interpret this behavior as follows.  From Eqs. (\ref{piadv}), (\ref{muadv}), (\ref{numu}), (\ref{numubar}) and (\ref{nue}), in the limit of $t_{\rm acc} \ll \tau_{\pi},\tau_{\mu}$ and $t_{\rm dyn} \gg \tau_{\pi},\tau_{\mu}$, the neutrino fluxes from advected particles can be approximated as
\begin{eqnarray}
\Phi_{\nu_e,{\rm nonacc}}^0(p)&\simeq &V\cdot 4\pi p^2 \xi_{\mu}^{-1}Q_{\pi,0}(p/\xi_{\mu}\xi_{\nu_e}), \label{nuenonacclow}\\
\Phi_{\nu_{\mu},{\rm nonacc}}^0(p)+\Phi_{\nu_{\bar{\nu}_{\mu}},{\rm nonacc}}^0(p)&\simeq& V\cdot 4\pi p^2 \left[ Q_{\pi,0}(p/\xi_{\nu_{\mu}}) + \xi_{\mu}^{-1}Q_{\pi,0}(p/\xi_{\mu}\xi_{\bar{\nu}_{\mu}}) \right], \label{numunonacclow}
\end{eqnarray} 
while in the limit of $t_{\rm acc} \ll \tau_{\pi},\tau_{\mu}$ and $t_{\rm dyn} \ll \tau_{\pi},\tau_{\mu}$ they can be approximated as
\begin{eqnarray}
\Phi_{\nu_e,{\rm nonacc}}^0(p)&\simeq&V\cdot 4\pi p^2 \xi_{\mu}^{-1}(t_{\rm dyn}/\tau_{\mu})Q_{\pi,0}(p/\xi_{\mu}\xi_{\nu_e}), \label{nuenonacchigh}\\
\Phi_{\nu_{\mu},{\rm nonacc}}^0(p)+\Phi_{{\bar{\nu}_{\mu}},{\rm nonacc}}^0(p)&\simeq & V\cdot 4\pi p^2 t_{\rm dyn}\left[ \tau_{\pi}^{-1}Q_{\pi,0}(p/\xi_{\nu_{\mu}})+(\xi_{\mu}\tau_{\mu})^{-1}Q_{\pi,0}(p/\xi_{\mu}\xi_{\bar{\nu}_{\mu}}) \right], \label{numunonacchigh}
\end{eqnarray}
where $V$ is the volume of the merged shell making the internal shock.  Here we neglect the contribution from pions/muons in the upstream region ($f_{\pi / \mu,-}(x,p)$) because it is subdominant compared to that from the downstream pions/muons.  We can easily see that in the latter limit $t_{\rm dyn} \ll \tau_{\pi}, \tau_{\mu}$, the energy spectra of neutrino fluxes are softer than $Q_{\pi,0}$ by $p^1$ because the decay timescale $\tau_i$ is proportional to $p$.  

On the other hand, the neutrino fluxes from reaccelerated pions/muons increase more as the acceleration timescales become longer.  Under the condition $t_{\rm acc} \ll \tau_{\pi}, \tau_{\mu}$, from Eqs. (\ref{piacc}), (\ref{muacc}), (\ref{numu}), (\ref{numubar}) and (\ref{nue}), we can approximate the neutrino fluxes from reaccelerated pions/muons as
\begin{eqnarray}
\Phi_{\nu_e,{\rm acc}}^0(p)&\simeq& S u_+ 4\pi p^2 \xi_{\nu_e}^{-1}f_{\mu,0}(p/\xi_{\nu_e}) \left\{ 1- \exp \left( -\frac{\xi_{\nu_e}r_d/\Gamma}{u_+\tau_{\mu}} \right) \right\}, \\
\Phi_{\nu_{\mu},{\rm acc}}^0(p)+\Phi_{\bar{\nu}_{\mu},{\rm acc}}^0(p)&\simeq& S u_+ 4\pi p^2 \left[ \xi_{\nu_{\mu}}^{-1}f_{\pi,0}(p/\xi_{\nu_{\mu}}) \left\{ 1 - \exp \left( -\frac{\xi_{\nu_{\mu}}r_d/\Gamma}{u_+\tau_{\pi}} \right) \right\}  \right. \nonumber \\
&&\left. + \xi_{\bar{\nu}_{\mu}}^{-1} f_{\mu,0}(p/\xi_{\bar{\nu}_{\mu}}) \left\{ 1 -\exp \left( -\frac{\xi_{\bar{\nu}_{\mu}}r_d/\Gamma}{u_+\tau_{\mu}} \right) \right\} \right].
\end{eqnarray}
In the high energy limit, where the decay timescales of pions/muons are much longer than the dynamical timescale, each of these neutrino fluxes behaves asymptotically as
\begin{eqnarray}
\Phi_{\nu_e}^0 &\sim &V\cdot \frac{4\pi p^2 f_{\mu,0}(p/\xi_{\nu_e})}{\tau_{\mu}} \propto p^2 Q_{\pi,0} \frac{t_{\rm acc}^2}{\tau_{\pi}\tau_{\mu}}, \label{nuehelimit}\\
\Phi_{\nu_{\mu}}^0 &\sim &V\cdot \frac{4\pi p^2 f_{\pi,0}(p/\xi_{\nu_{\mu}})}{\tau_{\pi}} \propto p^2 Q_{\pi,0} \frac{t_{\rm acc}}{\tau_{\pi}} \label{numuhelimit}, \\
\Phi_{\bar{\nu}_{\mu}}^0 &\sim &V\cdot \frac{4\pi p^2 f_{\mu,0}(p/\xi_{\bar{
\nu}_{\mu}})}{\tau_{\mu}} \propto p^2 Q_{\pi,0} \frac{t_{\rm acc}^2}{\tau_{\pi}\tau_{\mu}} \label{numubarhelimit}, 
\end{eqnarray}
where we use the definition $t_{\rm acc}=D(p)/u_-^2$ and the approximate expressions, Eqs. (\ref{fpizeroapp}) and (\ref{fmuzeroapp}).

Figure 4 depicts the neutrino flavor ratios as functions of energy expected from the internal shock of ultralong GRBs occurring inside progenitors.  In addition to the plot for the parameter set used in the previous figures (solid line), we show the ratio in the case with longer acceleration timescale for comparison (dashed line).  In the usual case, the flavor ratio expected from the photomeson process is $\Phi_{\nu_e}^0:\Phi_{\nu_{\mu}}^0:\Phi_{\nu_{\tau}}^0=1:2:0$, being independent of energy.  However, when the decay timescale of a muon becomes longer than the dynamical timescale, the flavor ratio is modified because the decay timescale of a muon is $\sim 100$ times longer than that of a pion and only the $\nu_e$ flux is reduced.  On the other hand, the acceleration of pions and muons also modifies the flavor ratio, and dominates the neutrino fluxes when the acceleration timescale becomes comparable to the dynamical timescale.  The flavor ratio becomes constant in the high energy limit.  We can explain this behavior from Eqs. (\ref{fpizeroapp}), (\ref{fmuzeroapp}), (\ref{nuehelimit}), (\ref{numuhelimit}) and (\ref{numubarhelimit}): the ratio $\Phi_{\nu_{\mu}}^0/\Phi_{\nu_e}^0$ is determined only by the ratio between the acceleration timescale $t_{\rm acc}$ and the decay timescale of a muon $\tau_{\mu}$, which is independent of momentum $p$.  More explicitly, when assuming a strong shock ($\sigma=4$, $\gamma=4$) with $Q_{\pi,0}(p)$ being proportional to $p^{-4}$ (i.e., $\alpha=4$), we can describe the flavor ratio at the source in the high energy limit as
\begin{eqnarray}
\frac{\Phi_{\nu_{\mu}}^0+\Phi_{\bar{\nu}_{\mu}}^0}{\Phi_{\nu_e}^0}&\simeq&\frac{\xi_{\nu_e}^{\alpha-2}\xi_{\mu}^{\alpha-1}\left[ \left( \frac{1}{\xi_{\mu}}+\sigma \right) \left( \frac{1}{\xi_{\pi}}+\sigma \right) \frac{\gamma}{\gamma-\alpha+1}+\frac{1}{\xi_{\mu}\xi_{\pi}^2} \right] }{\xi_{\nu_{\mu}}^{\alpha-1}\left( \frac{1}{\xi_{\pi}}+\sigma \right) }\frac{\tau_{\mu}}{t_{\rm acc}}+1 \nonumber \\
&\simeq& 0.022\frac{\tau_{\mu}}{t_{\rm acc}}+1. \label{ratiosource}
\end{eqnarray}
This ratio diverges in the limit of $\tau_{\mu}/t_{\rm acc} \rightarrow \infty$, which means that the flavor ratio at the source, $\Phi_{\nu_e}^0:\Phi_{\nu_{\mu}}^0:\Phi_{\nu_{\tau}}^0$ approaches $0:1:0$.  Interestingly, we may be able to infer the particle-acceleration timescale from the neutrino flavor ratio.

By using Eq. (\ref{mixing}), we can evaluate the neutrino flavor ratio that would be observed at the Earth, as shown in Figure 5.  Similar to the flavor ratio at the source, the observed ratio is modified above the energy where the decay timescale of a muon becomes longer than the dynamical timescale and is nearly constant in the high energy range.  The flavor transition occurs over $\sim 2$ decades in energy.

We can easily show that, in the limit of $\tau_{\mu}/t_{\rm acc} \rightarrow \infty$, the observed flavor ratio $\Phi_{\nu_e}:\Phi_{\nu_{\mu}}:\Phi_{\nu_{\tau}}$ at high energy range converges to $\simeq 1:1.8:1.8$.  This ratio is identical to that shown in \cite{kashtiwaxman05}, in which they investigated the effects of synchrotron cooling of pions/muons before their decay on the neutrino flavor ratio.  In their work, as in our current study, the neutrino flavor ratio at the source is $0:1:0$ at high energy, but the reason is different.  In the model of \cite{kashtiwaxman05}, since the lifetime of a muon is longer than that of a pion, muons would suffer from synchrotron cooling more than pions.  As a result, the flux of electron neutrinos, that are produced from muons, would be suppressed compared to the flux of muon neutrinos, that are produced from pions.  Therefore, in the high energy range where the synchrotron cooling timescale is much shorter than the lifetime of a muon, the flavor ratio at the source can be approximated as $\simeq 0:1:0$.  In our study, we show that the flavor ratio would be also modified by the secondary-acceleration because pions decay more than muons during the secondary-acceleration, if the acceleration timescale of a pion/muon is shorter than their lifetimes.

On the contrary to the cooling case in which the flavor modification is associated with the spectral softening, the neutrino spectra are flat in the high energy range when pions and muons are reaccelerated.  This is because the secondary-acceleration makes the spectra of primary pions/muons harder by $p^1$ [$\propto D(p)$ in Eqs,~(\ref{fpizeroapp}) and (\ref{fmuzeroapp})], while the neutrino spectra is softer than the primary spectra by $p^1$, which is proportional to the decay timescales of primary particles.  As a result, the neutrino spectra are flat, having the same spectral indices with those of injected primary mesons.  Therefore, even when the observed flavor ratio of neutrinos in the high energy range converges to $1:1.8:1.8$, we can discriminate which process modifies the flavor ratio, cooling or secondary-acceleration, by observing their energy spectra.  We should note that the flat part of neutrino spectra would have a cutoff at the energy where the acceleration timescale is equal to the dynamical timescale because above that energy pions and muons would suffer from adiabatic cooling, which is not included in our formulation (see discussion in Sec. 4).

\section{Discussion and Conclusion}

We investigate the shock acceleration of pions/muons produced by primary protons that are accelerated at the shock, and its effects on the observed neutrino flavor ratios.  We solve the convection-diffusion equation of pions/muons around a shock taking  secondary-acceleration and decay into account, and compute the high energy neutrino spectra from their decay as well as the energy dependence of the neutrino flavor ratio $\Phi_{\nu_e}:\Phi_{\nu_{\mu}}:\Phi_{\nu_{\tau}}$.  We find the following:

1. When the acceleration timescale is shorter than the decay timescales of a pion and a muon, pions and muons are accelerated at the shock before they decay.  The resulting distribution function of pions/muons would be divided into two components: the component accelerated at the shock and the component advected to the downstream after production from protons.  The neutrino spectrum of the former component is flat in the high energy range where the acceleration timescale becomes comparable to the dynamical timescale of the system.

2.  The flavor ratio of neutrinos at the source, $\Phi_{\nu_e}^0:\Phi_{\nu_{\mu}}^0:\Phi_{\nu_{\tau}}^0$, would deviate from $1:2:0$, which is expected from photomeson interactions, and approaches to $0:1:0$ above the energy at which the decay timescale of a muon becomes longer than the dynamical timescale of the shock because only the $\nu_e$ flux is reduced.  The transition width of the observed flavor ratio is $\sim 2$ decades in energy (Fig. 5), which is wider than that in the case of the flavor ratio modification by the radiative cooling.  Although such a flavor ratio modification by the adiabatic cooling has been suggested in \cite{kashtiwaxman05}, we investigate them using the convection-diffusion equation for the first time.

3. When the secondary-acceleration is efficient, the neutrino fluxes from shock-reaccelerated pions/muons are dominant over the fluxes from non-reaccelerated pions/muons in the high energy range.  In this case the flavor ratio would be asymptotically constant (Fig.4).  This ratio is determined by the ratio of the lifetime of a muon to its acceleration timescale (see Eq. \ref{ratiosource}).  Therefore, from the observed flavor ratio, one can constrain the acceleration timescale of cosmic ray particles.

4. The maximum energy of accelerated particles is determined by the condition $t_{\rm acc}=t_{\rm dyn}$, where the energy spectra of neutrinos become flat.  As a result, the secondary-accelerated component appears as a flat excess above the non-reaccelerated component at the highest energy.

5. When the acceleration timescale is shorter than the lifetime of a muon, the flavor ratio at the source approaches to $\Phi_{\nu_e}^0:\Phi_{\nu_{\mu}}^0:\Phi_{\nu_{\tau}}^0 \rightarrow 0:1:0$ in the high energy range, and the observed flavor ratio approaches to $1:1.8:1.8$.  This asymptotic ratio is similar to the case where pions/muons are efficiently cooled via synchrotron and/or inverse Compton scattering, but the energy spectra of neutrinos are different: the spectra become flat in the high energy range when the secondary-acceleration is efficient, while the spectra become soft in the high energy range when the synchrotron cooling is efficient.

6.  As for the ratio of $\bar{\nu}_e$ to the total $\nu$ flux, when $t_{\rm acc} \ll \tau_{\pi}, \tau_{\mu}$, it is $\sim 1/14$ ($\sim 1/6$) in the low energy range and it approaches to $\sim 0$ ($\sim 1/9$) in the case of $p\gamma$ ($pp$) interactions.  The ratio of the $\bar{\nu}_e$ flux to the total $\nu$ flux in the high energy range can be measured from the $\bar{\nu}_e$ interactions at the $6.3~{\rm PeV}$ Glashow resonance.

Our formalism presented in Section II can be applied only when the flow speed at the shock rest frame is non-relativistic.  In the case when the shock is relativistic, we should use relativistic formulae for shock-acceleration, in which the anisotropies in the angular distribution of accelerated particles are taken into account \cite{achterberg+01}.  Recent particle-in-cell (PIC) simulations of relativistic shocks have shown that the efficiency of particle acceleration is controlled by the magnetization, flow velocity, and field direction \cite{sironispitkovsky09, sironispitkovsky11, sironi+13}, and one should take into account these properties when discussing secondary-acceleration in relativistic shocks.  These issues would be important in future work.

In the calculations above, we neglect the radiative cooling of pions and muons during the shock acceleration.  If the energy of pions or muons is higher than $\varepsilon_{i,0}$ in Eqs. (\ref{epizero}) and (\ref{emuzero}), we should consider the synchrotron cooling in deriving the distribution functions of pions and muons.  As shown in \cite{kashtiwaxman05}, due to the synchrotron cooling of pions and muons, the observed neutrino flavor ratio, $\Phi_{\nu_e}:\Phi_{\nu_{\mu}}:\Phi_{\nu_{\tau}}$, is modified from $1:1:1$ at low energy to $1:1.8:1.8$ at high energy.  In this energy range, the energy spectra of neutrinos are softened.  This expectation will be confirmed by solving the pion/muon transport equations with the energy loss term (e.g. \cite{blasi10}).  We also neglected the effect of matter oscillations (Mekheyev-Smirnov-Wolfenstein effect), which would be important in the case of neutrino emission from the GRB jet inside a star because of the high density \cite{waxmanbahcall97, fraija15, xiaodai15}.  These are interesting future works.

\ \\

We thank K. Kohri, K. Asano, R. Yamazaki, H. Takami, K. Murase and K. Kashiyama for useful comments.  This work is supported by the Grants-in-Aid for Scientific Research No. 26287051, 26287051, 24103006, 24000004 and 26247042 (K.I.).

\begin{figure}[htbp]
\includegraphics[scale=1.3]{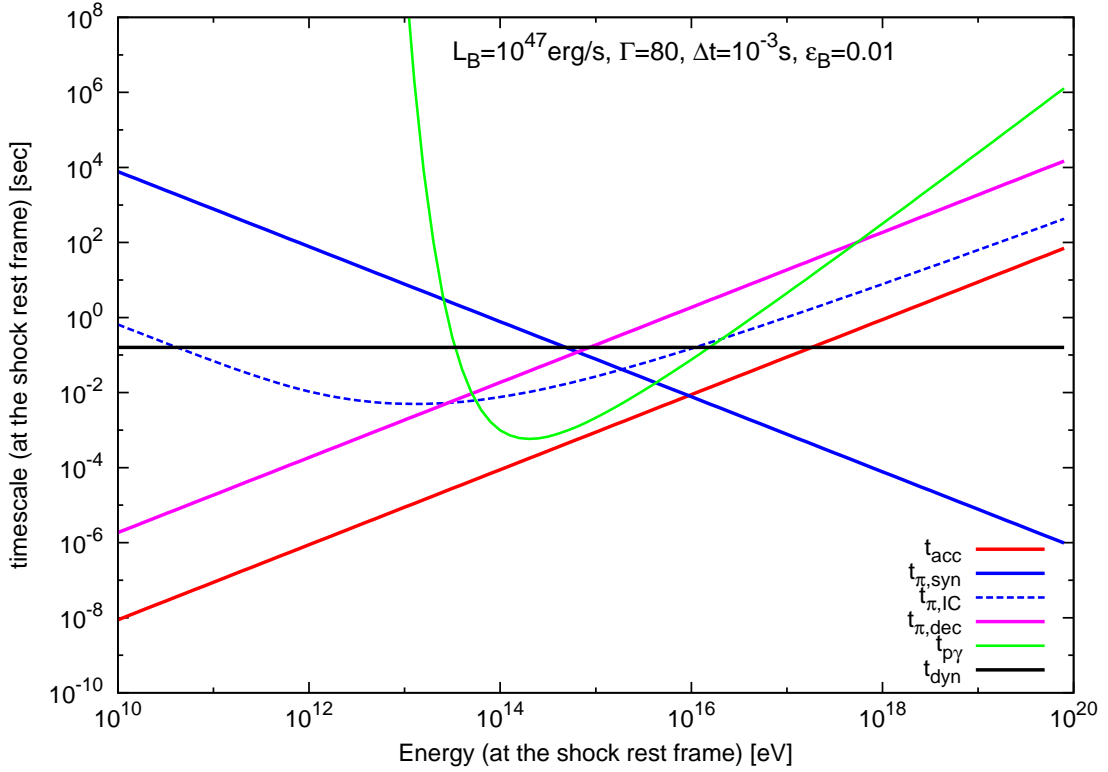}
\caption{The acceleration timescale, cooling timescales via synchrotron emission and inverse Compton scattering, and decay timescale of charged pions $\pi^+$ in the internal shock occurring inside a progenitor of an ultralong GRB (measured at the shock rest frame).  The photomeson timescale and dynamical timescale are also shown.  Used parameters are $L_B=10^{47}~{\rm erg}~{\rm s}^{-1}$, $\Gamma=80$, $\Delta t=10^{-3}~{\rm s}$, $\beta_-=0.5$, and $\epsilon_{B}=0.01$.
}
\label{f1pi}
\end{figure}

\begin{figure}[htbp]
\includegraphics[scale=1.3]{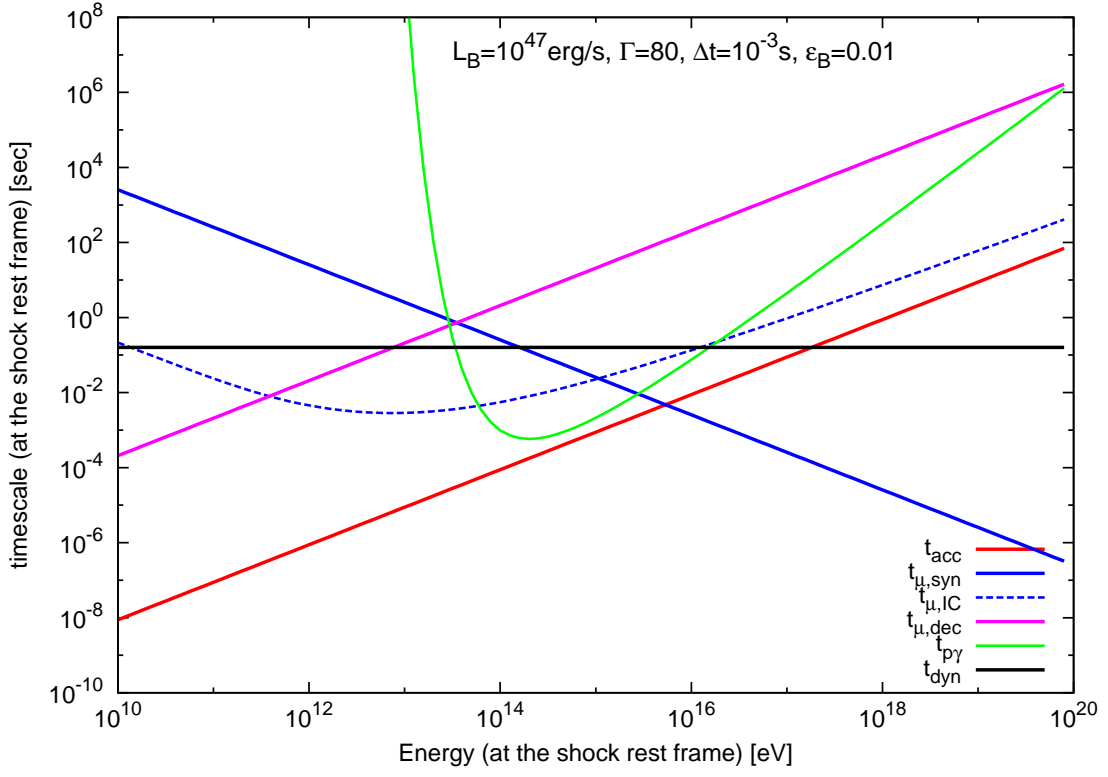}
\caption{The same as in Fig. 1, but for muons $\mu^{\pm}.$
}
\label{f1mu}
\end{figure}

\begin{figure}[htbp]
\includegraphics[scale=1.3]{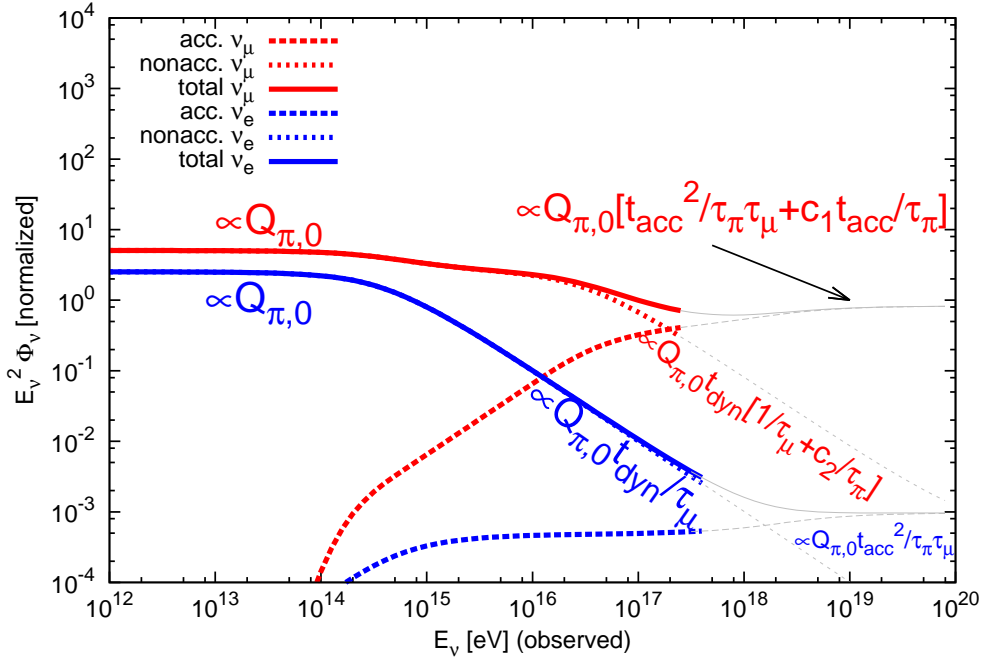}
\caption{The energy flux of $\nu_{\mu}+\bar{\nu}_{\mu}$ (red lines) and $\nu_e$ (blue lines) expected from a low-power GRB (where the flavor oscillation during propagation is not taken into account), normalized to the flux of electron neutrinos $E_{\nu}^2\Phi_{\nu_e}$ at low energy.  Used parameters are the same as in Fig. 1., and the pion production spectrum is assumed as $Q_{\pi,0}(p) \propto p^{-\alpha}$ where $\alpha=4$.  The muon neutrino flux and the electron neutrino flux are divided into two components: those coming from reaccelerated pions and/or muons (dashed lines) and those coming from the pions and/or muons advected to the downstream region (dotted lines).  In the low energy range, the $\nu_e$ flux and $\nu_{\mu}+\bar{\nu}_{\mu}$ flux are dominated by the latter component [$\propto Q_{\pi,0}$, see Eq.(\ref{nuenonacclow}) and (\ref{numunonacclow})].  The $\nu_e$ and $\nu_{\mu}+\bar{\nu}_{\mu}$ fluxes drop above the energy where the decay timescales of a muon and a pion are equal to the dynamical timescale, being proportional to $\simeq Q_{\pi,0}t_{\rm dyn}/\tau_{\mu}$ and  $\simeq Q_{\pi,0}t_{\rm dyn}[1/\tau_{\mu}+c_2/\tau_{\pi}]$, respectively, where $c_2\simeq \xi_{\mu}^{-\alpha+1}(\xi_{\nu_{\mu}}\xi_{\bar{\nu}_{\mu}})^{\alpha}\simeq 10^{-4}$ is a constant [see Eqs. (\ref{nuenonacchigh}) and (\ref{numunonacchigh})].  In the high energy range, the fluxes are dominated by the neutrinos from reaccelerated pions and/or muons: the $\nu_e$ and $\nu_{\mu}+\bar{\nu}_{\mu}$ fluxes are proportional to $\simeq Q_{\pi,0}t_{\rm acc}^2/\tau_{\pi}\tau_{\mu}$ and $\simeq Q_{\pi,0}[t_{\rm acc}^2/\tau_{\pi}\tau_{\mu}+c_1t_{\rm acc}/\tau_{\pi}]$, respectively, where $c_1$ is a constant, the coefficient in front of $\tau_{\mu}/t_{\rm acc}$ in Eq. (\ref{ratiosource}) [see Eqs. (\ref{nuehelimit}), (\ref{numuhelimit}) and (\ref{numubarhelimit})].  The sum of two components are shown by solid lines.   Note that,  if the energy is higher than $\sim 2\times 10^{16}~{\rm eV}$ at the observed frame ($\sim 3\times 10^{15}~{\rm eV}$ at the shock rest frame), the spectra would have a cutoff due to the synchrotron cooling of pions and muons (see Figs. 1 and 2), which is not taken into account in the current calculation, and therefore the plots above this energy (shown with thin grey lines) would be suppressed.  Note also that, if the energy is higher than a few times $\sim 10^{17}~{\rm eV}$ at the observed frame (a few times $10^{16}~{\rm eV}$ at the shock rest frame), where the acceleration timescale is longer than the dynamical timescale, the neutrino spectra would have a cutoff because there would be no accelerated particles generating neutrinos with such energy.}
\label{f3}
\end{figure}

\begin{figure}[htbp]
\begin{tabular}{cc}
\resizebox{80mm}{!}{\includegraphics{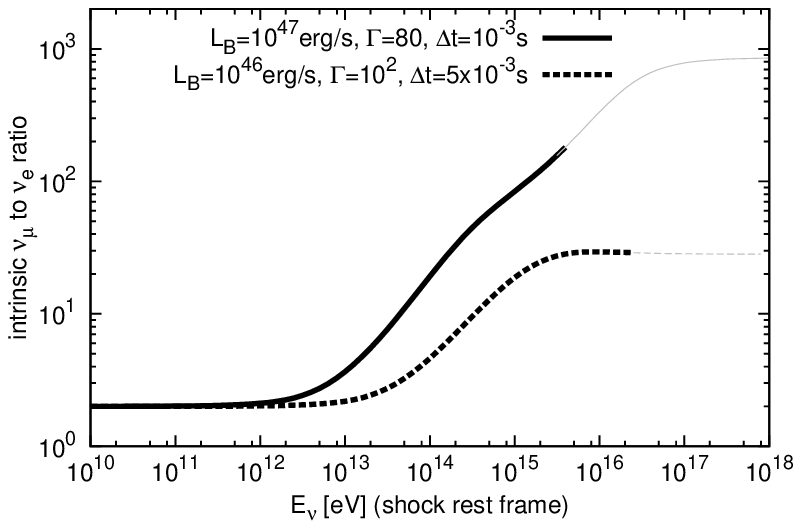}} &
\resizebox{80mm}{!}{\includegraphics{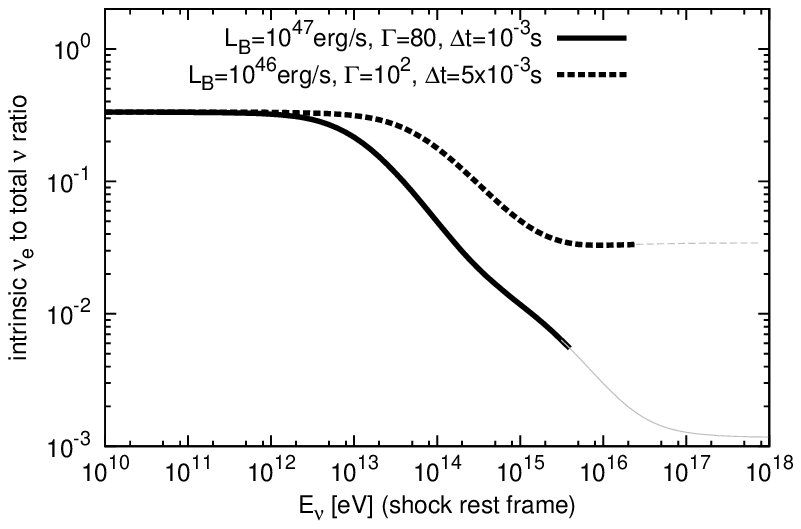}} \\
\end{tabular}
\caption{Energy dependence of the flavor ratio of $\nu_{\mu}+\bar{\nu}_{\mu}$ to $\nu_e$ ({\it left}) and that of the ratio of $\nu_e$ to the total neutrino flux ({\it right}) at the source for low-power GRBs.  Used parameter sets $(L_B, \Gamma, \Delta t, \beta_-)$ are $(10^{47}~{\rm erg}~{\rm s}^{-1}, 80, 10^{-3}~{\rm s}, 0.5)$ (solid line) and $(10^{46}~{\rm erg}~{\rm s}^{-1}, 10^2, 5\times10^{-3}~{\rm s}, 0.5)$ (dashed line).  As stated in the caption of Fig. 3, in the higher energy range where the acceleration timescale is longer than the cooling timescale and/or the dynamical timescale (shown with thin grey lines) the ratios would be modified from those shown in these figures.}
\label{f4}
\end{figure}

\begin{figure}[htbp]
\begin{tabular}{cc}
\resizebox{80mm}{!}{\includegraphics{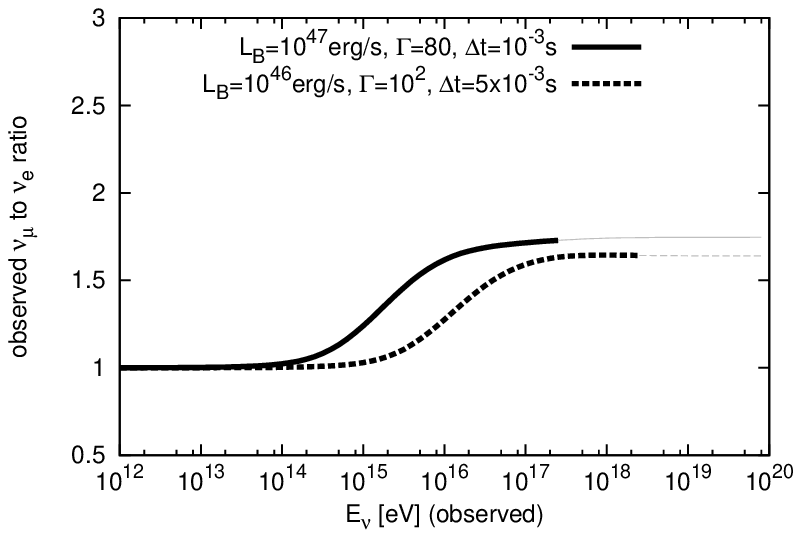}} &
\resizebox{80mm}{!}{\includegraphics{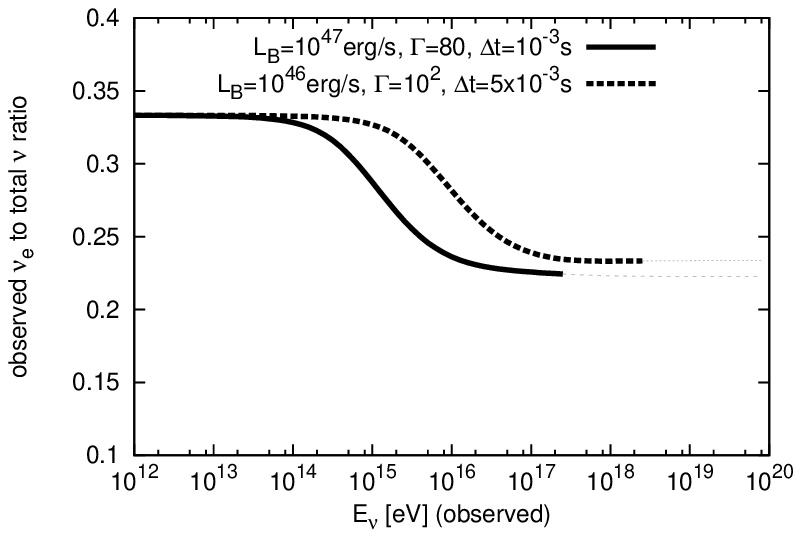}} \\
\end{tabular}
\caption{Energy dependence of the flavor ratio of $\nu_{\mu}+\bar{\nu}_{\mu}$ to $\nu_e+\bar{\nu}_e$ ({\it left}) and that of the ratio of $\nu_e+\bar{\nu}_e$ to the total neutrino flux ({\it right}) observed at the Earth for low-power GRBs.  Used parameter sets $(L_B, \Gamma, \Delta t, \beta_-)$ are $(10^{47}~{\rm erg}~{\rm s}^{-1}, 80, 10^{-3}~{\rm s}, 0.5)$ (solid line) and $(10^{46}~{\rm erg}~{\rm s}^{-1}, 10^2, 5\times10^{-3}~{\rm s}, 0.5)$ (dashed line).  Flavor oscillation during propagation is taken into account.  As in Figs. 3 and 4, in the high energy range (thin grey lines) the ratios would be modified.}
\label{f5}
\end{figure}


%


\appendix
\section{General Solutions of the Convection-Diffusion Equation for Decaying Particles}

In this section we discuss how to solve the shock acceleration of charged particles decaying in a finite time such as pions and muons.  Their convection-diffusion equation is shown in Eq. (\ref{transport}).  The general solution can be described as
\begin{eqnarray}
f_i(x,p)=\left \{
\begin{array}{ll}
\int_{-\infty}^0 dx^{\prime}Q_i(x^{\prime},p)G_{i,-}(x^{\prime};x,p)+H_{i,-}(x,p) & (x \leq 0), \\
\int_0^{\infty}dx^{\prime}Q_i(x^{\prime},p)G_{i,+}(x^{\prime};x,p)+H_{i,+}(x,p) & (x>0), \\
\end{array} \right. \label{solution}
\end{eqnarray}
where $G_{i,\pm}(x^{\prime};x,p)$ are the Green functions of Eq.(\ref{transport}) with respect of $x$, and $H_{i,\pm}(x,p)$ are the homogeneous solutions of Eq. (\ref{transport}) which should be determined by the boundary conditions.

The Green functions of Eq.(\ref{transport}) are given by
\begin{eqnarray}
G_{i,\pm}(x^{\prime};x,p)=\left \{
\begin{array}{ll}
\frac{1}{\sqrt{u_{\pm}^2+4D/\tau_i}}\exp \left[ -\frac{\sqrt{u_{\pm}^2+4D/\tau_i}-u_{\pm}}{2D}(x-x^{\prime}) \right] & (x>x^{\prime}), \\
\frac{1}{\sqrt{u_{\pm}^2+4D/\tau_i}}\exp \left[ \frac{\sqrt{u_{\pm}^2+4D/\tau_i}+u_{\pm}}{2D}(x-x^{\prime}) \right] & (x<x^{\prime}), \\ 
\end{array} \right. \label{gipm}
\end{eqnarray}
and, under the condition (\ref{cd1}), the homogeneous solutions should be
\begin{eqnarray}
H_{i,\pm}(x,p)=\left \{
\begin{array}{ll}
\left[ f_{i,0}(p)-\frac{1}{\sqrt{u_-^2+4D/\tau_i}}\int_{-\infty}^0 dx^{\prime} Q_i(x^{\prime},p)\exp \left( \frac{\sqrt{u_-^2+4D/\tau_i}-u_-}{2D}x^{\prime} \right) \right]\exp \left( \frac{\sqrt{u_-^2+4D/\tau_i}+u_-}{2D}x \right) & (x \leq 0), \\
\left[ f_{i,0}(p)-\frac{1}{\sqrt{u_+^2+4D/\tau_i}}\int_0^{\infty} dx^{\prime} Q_i(x^{\prime},p) \exp \left( -\frac{\sqrt{u_+^2+4D/\tau_i}+u_+}{2D}x^{\prime} \right) \right]\exp \left( -\frac{\sqrt{u_+^2+4D/\tau_i}-u_+}{2D}x \right) & (x > 0). \\
\end{array} \right. \label{hipm}
\end{eqnarray}

The differential equation for $f_{i,0}(p)$ with respect of $p$ is given by the condition (iii) in Eq. (\ref{cd3}), which can be rewritten as
\begin{eqnarray}
p\frac{\partial f_{i,0}}{\partial p}=-\gamma A_i f_{i,0}+\gamma g_i(p), \label{fizero}
\end{eqnarray}
where $\gamma=3\sigma/(\sigma-1)$ ($\sigma=u_-/u_+$ is the compression ratio), $A_i$ is the numerical factor, being independent of $p$ (since both $D$ and $\tau_{\mu}$ are proportional to $p$):
\begin{eqnarray}
A_i=\frac{1}{2} \left[ \left( \sqrt{1+\frac{4D}{\tau_i u_-^2}}+1 \right) + \left( \sqrt{\frac{1}{\sigma^2}+\frac{4D}{\tau_i u_-^2}}-\frac{1}{\sigma} \right) \right], \label{ai}
\end{eqnarray}
and $g_i(p)$ is given by
\begin{eqnarray}
g_i(p)=\frac{1}{u_-}\left[ \int_{-\infty}^0 dx^{\prime} Q_i(x^{\prime},p)\exp \left( \frac{ \sqrt{u_-^2+4D/\tau_i}-u_-}{2D}x^{\prime} \right) + \int_0^{\infty} dx^{\prime} Q_i(x^{\prime},p)\exp \left( -\frac{ \sqrt{u_+^2+4D/\tau_i}+u_+}{2D}x^{\prime} \right) \right]. \label{gip}
\end{eqnarray}
One can generally solve Eq.(\ref{fizero}) as
\begin{eqnarray}
f_{i,0}(p)=\int_0^p \frac{dp^{\prime}}{p^{\prime}} \left( \frac{p^{\prime}}{p} \right)^{\gamma A_i} g_i(p^{\prime}). \label{fizeroint}
\end{eqnarray} 

\section{Solution of the Convection-Diffusion Equation for Muons}
In this section we describe the solution of the convection-diffusion equation for muons (decaying tertiary particles) in detail.  The production spectrum of muons at the source (per unit time, per unit spatial volume, and per unit volume of the momentum space) should be given based on the distribution function of pions as follows:
\begin{eqnarray}
Q_{\mu,-}&=&\frac{1}{\xi_{\mu}}\frac{f_{\pi,-}(p/\xi_{\mu})}{\tau_{\pi}(p/\xi_{\mu})}=q_{\mu,{\rm a}}^{-}\exp \left( \frac{\sqrt{u_-^2+4D/\tau_{\pi}}+u_-}{2D/\xi_{\mu}}x \right) +q_{\mu,{\rm b}}^{-}\exp \left(\frac{\xi_{\pi}\xi_{\mu}u_-}{D}x \right), \\
Q_{\mu,+}&=&\frac{1}{\xi_{\mu}}\frac{f_{\pi,+}(p/\xi_{\mu})}{\tau_{\pi}(p/\xi_{\mu})}=q_{\mu,{\rm a}}^{+}\exp \left( -\frac{\sqrt{u_+^2+4D/\tau_{\pi}}-u_+}{2D/\xi_{\mu}}x \right) +q_{\mu,{\rm b}}^{+},
\end{eqnarray}
where $\xi_{\mu}\approx 0.75$ is the ratio of the energy of a muon to that of a primary pion, and the functions $q_{\mu, {\rm a}}^{\pm}(p)$ and $q_{\mu,{\rm b}}^{\pm}(p)$ are given as
\begin{eqnarray}
q_{\mu,{\rm a}}^-(p)&=&\frac{1}{\tau_{\pi}} \left[ f_{\pi,0}(p/\xi_{\mu})-\frac{D(p)Q_{\pi,0}(p/\xi_{\mu})}{\xi_{\mu}\left( D/\tau_{\pi}+(\xi_{\pi}-\xi_{\pi}^2)u_-^2 \right) } \right], \\
q_{\mu,{\rm b}}^-(p)&=&\frac{1}{\tau_{\pi}} \left[ \frac{D(p)Q_{\pi,0}(p/\xi_{\mu})}{\xi_{\mu}\left( D/\tau_{\pi}+(\xi_{\pi}-\xi_{\pi}^2)u_-^2 \right) } \right], \\
q_{\mu,{\rm a}}^+(p)&=&\frac{f_{\pi,0}(p/\xi_{\mu})}{\tau_{\pi}} -\frac{1}{\xi_{\mu}}Q_{\pi,0}(p/\xi_{\mu}), \\
q_{\mu,{\rm b}}^+(p)&=&\frac{1}{\xi_{\mu}}Q_{\pi,0}(p/\xi_{\mu}).
\end{eqnarray}
Note that, since $\tau_{\mu}(p)$ is approximately proportional to $p$, the momentum dependence of $Q_{\mu}(p)$ is softer than $f_{\pi}(p)$.

Substituting this $Q_{\mu}(x,p)$, we can obtain the muon distribution function in the upstream $f_{\mu,-}(x,p)$ and downstream $f_{\mu,+}(x,p)$ as
\begin{eqnarray}
f_{\mu,-}&=& \left( f_{\mu,0}-\frac{4Dq_{\mu,{\rm a}}^{-}}{(u_-^2+4D/\tau_{\mu})-(\xi_{\mu}\sqrt{u_-^2+4D/\tau_{\pi}}-(1-\xi_{\mu})u_-)^2} -\frac{Dq_{\mu,{\rm b}}^{-}}{D/\tau_{\pi}+(\xi_{\mu}\xi_{\pi}-\xi_{\mu}^2 \xi_{\pi}^2)u_-^2}\right)  \nonumber\\
&&\times \exp \left( \frac{\sqrt{u_-^2+4D/\tau_{\mu}}+u_-}{2D}x \right) \nonumber \\
&&+\frac{4Dq_{\mu,{\rm a}}^-}{(u_-^2+4D/\tau_{\mu})-(\xi_{\mu}\sqrt{u_-^2+4D/\tau_{\pi}}-(1-\xi_{\mu})u_-)^2} \exp \left( \frac{\sqrt{u_-^2+4D/\tau_{\pi}}+u_-}{2D/\xi_{\mu}}x \right) \nonumber \\
&&+\frac{Dq_{\mu,{\rm b}}^-}{D/\tau_{\pi}+(\xi_{\mu}\xi_{\pi}-\xi_{\mu}^2 \xi_{\pi}^2)u_-^2}\exp \left( \frac{\xi_{\mu} \xi_{\pi}u_-}{D}x \right), \\
f_{\mu,+}&=&\left(f_{\mu,0}-\frac{4Dq_{\mu,{\rm a}}^+}{(u_+^2+4D/\tau_{\mu})-(\xi_{\mu}\sqrt{u_+^2+4D/\tau_{\pi}}+(1-\xi_{\mu})u_+)^2} -q_{\mu,{\rm b}}^+ \tau_{\mu} \right) \exp \left(-\frac{\sqrt{u_+^2+4D/\tau_{\mu}}-u_+}{2D}x\right) \nonumber \\
&&+\frac{4Dq_{\mu,{\rm a}}^+}{(u_+^2+4D/\tau_{\mu})-(\xi_{\mu}\sqrt{u_+^2+4D/\tau_{\pi}}+(1-\xi_{\mu})u_+)^2} \exp\left( -\frac{\sqrt{u_+^2+4D/\tau_{\pi}}-u_+}{2D/\xi_{\mu}}x \right) +q_{\mu,{\rm b}}^+ \tau_{\mu}.
\end{eqnarray}

At the shock front, the muon distribution function should satisfy
\begin{eqnarray}
p\frac{df_{\mu,0}}{dp}&=&-\gamma A_{\mu}f_{\mu,0} +\frac{\gamma D(p)}{u_-^2}\left( q_{\mu,{\rm a}}^- B_{\mu,{\rm a}}^- +q_{\mu,{\rm b}}^- B_{\mu,{\rm b}}^- +q_{\mu,{\rm a}}^+ B_{\mu,{\rm a}}^+ +q_{\mu,{\rm b}}^+ B_{\mu,{\rm b}}^+ \right), \label{muonsf}
\end{eqnarray}
where $A_{\mu}$, $B_{\mu,{\rm a}}^{\pm}$ and $B_{\mu,{\rm b}}^{\pm}$ are numerical factors, being independent of $p$:
\begin{eqnarray}
A_{\mu}&=&\frac{1}{2}\left[ \left( \sqrt{1+\frac{4D}{\tau_{\mu}u_-^2}}+1\right)+\left( \sqrt{\frac{1}{\sigma^2}+\frac{4D}{\tau_{\mu}u_-^2}}-\frac{1}{\sigma}\right) \right], \\
B_{\mu,{\rm a}}^-&=&\frac{2}{\sqrt{1+4D/\tau_{\mu}u_-^2}-1+\xi_{\mu}\left( \sqrt{1+4D/\tau_{\pi}u_-^2}+1\right) }, \\
B_{\mu,{\rm b}}^-&=&\frac{2}{\sqrt{1+4D/\tau_{\mu}u_-^2}-(1-2\xi_{\mu}\xi_{\pi})}, \\
B_{\mu,{\rm a}}^+&=&\frac{2\sigma}{\sqrt{1+4D/\tau_{\mu}u_+^2}+1+\xi_{\mu}\left( \sqrt{1+4D/\tau_{\pi}u_+^2}-1 \right) }, \\
B_{\mu,{\rm b}}^+&=&\frac{2\sigma}{\sqrt{1+4D/\tau_{\mu}u_+^2}+1}.
\end{eqnarray}

One can solve Eq.(\ref{muonsf}) as
\begin{eqnarray}
f_{\mu,0}(p)&=&\gamma\int_0^p \frac{dp^{\prime}}{p^{\prime}} \left( \frac{p^{\prime}}{p} \right) ^{\gamma A_{\mu}}\frac{D(p^{\prime})}{u_-^2} \left( q_{\mu,{\rm a}}^- (p^{\prime})B_{\mu,{\rm a}}^-  +q_{\mu,{\rm b}}^-(p^{\prime}) B_{\mu,{\rm b}}^- +q_{\mu,{\rm a}}^+ (p^{\prime})B_{\mu,{\rm a}}^+ +q_{\mu,{\rm b}}^+ (p^{\prime})B_{\mu,{\rm b}}^+ \right).
\end{eqnarray}




\begin{thebibliography}{99}
\bibitem{muonadcool}
  The muons whose decay timescale is longer than the dynamical timescale would suffer from adiabatic cooling and lose their energy as $\dot{\varepsilon}_{\mu} \propto -\varepsilon^1_{\mu}$, and the resulting energy spectra of neutrinos would be softened by a factor of $\varepsilon_{\nu}^1$.
\end{thebibliography}
\end{document}